\documentclass[pdftex,twocolumn,epjc3]{svjour3}          % twocolumn

\RequirePackage[T1]{fontenc}
\usepackage[utf8]{inputenc}
\DeclareUnicodeCharacter{2212}{-}
\usepackage{amsmath}     % for math environments
\usepackage{amssymb}     % for symbols
\usepackage{graphicx}    % for figures
\usepackage{textcomp}    % for text symbols

\smartqed  % flush right qed marks, e.g. at end of proof
\usepackage{physics}
\RequirePackage{graphicx}
\RequirePackage{mathptmx}      % use Times fonts if available on your TeX system
\RequirePackage{flushend}
\RequirePackage[numbers,sort&compress]{natbib}
\RequirePackage[colorlinks,citecolor=blue,urlcolor=blue,linkcolor=blue]{hyperref}
\usepackage{amsmath}
\usepackage{amssymb}
\usepackage{url}
\usepackage{subfig}  % or subcaption (but subfig is more EPJC-compatible)

\labelformat{section}{Section #1} 
\labelformat{subsection}{Section #1} 
\labelformat{subsubsection}{Section #1}
\labelformat{subsubsubsection}{Section #1}
\labelformat{equation}{Eq.~(#1)} 
\labelformat{figure}{Fig.~[#1]} 
\labelformat{subfigure}{Fig.~[\thefigure#1]} 
\labelformat{table}{Tab.~#1} 
\labelformat{appendix}{Appendix #1}

\journalname{Eur. Phys. J. C}

\begin{document}

\title{Radiative and Jet Signatures of Regular Black Holes in Quantum-Corrected Gravity}

\subtitle{Constraints from Disk Efficiency and Jet Power in Asymptotically Safe Spacetime}

\author{Chirantana Bhattacharjee\thanksref{e1,addr1,addr3}
     \and
        Subhadip Sau\thanksref{e2,addr2} \and Avijit Mukherjee\thanksref{e3,addr3}%etc.
}

%\thankstext[$\star$]{t1}{Thanks to the title}
\thankstext{e1}{e-mail: chirantana.physics@gmail.com}
\thankstext{e2}{e-mail: subhadipsau2@gmail.com}
\thankstext{e3}{e-mail: avijit.mukherjee@jadavpuruniversity.in}

\institute{Department of Physics, Asutosh College, Kolkata - 700026\label{addr1}
          \and
          Department of Physics, Jhargram Raj College, Jhargram, West Bengal - 721507\label{addr2}
          \and
        Department of Physics, Jadavpur University, Kolkata- 700032\label{addr3}
}

\date{Received: date / Accepted: date}

\maketitle

\begin{abstract}
	We investigate the observational viability of regular rotating black holes emerging from asymptotically safe gravity, a quantum gravitational framework where spacetime curvature is modified through a scale-dependent Newton's constant. By incorporating ultraviolet corrections to the near-horizon geometry, these solutions deviate from the classical Kerr metric while preserving asymptotic flatness and avoiding central singularities. In such spacetimes, both the radiative efficiency of accretion disks and the power output of relativistic jets are sensitive to the deformation parameter governing the quantum corrections.
	
	We compute the theoretical predictions for radiative efficiency and Blandford–Znajek jet power in quantum corrected rotating geometries and compare them with observational estimates for six well-studied stellar mass black holes. Our analysis reveals that for several systems with low to moderate spin, the asymptotically safe regular black hole model successfully reproduces both observables within reported uncertainties. In contrast, highly spinning systems such as GRS~1915$+$105 challenge the compatibility of this framework, suggesting a restricted deformation range or the need for additional physical inputs.
	
	The results demonstrate that quantum corrections, although confined to the strong field regime, can leave measurable imprints on high-energy astrophysical processes. Radiative and jet-based diagnostics thus serve as complementary probes of near-horizon geometry and provide a novel pathway to test quantum gravitational effects using electromagnetic observations. This work illustrates how precision measurements of spin, luminosity, and jet dynamics can offer indirect access to the ultraviolet structure of spacetime, motivating future studies of gravitational wave signatures, polarization spectra, and photon ring morphology in the presence of scale-dependent gravity.
\end{abstract}

\section{Introduction}

Penrose’s singularity theorem \cite{PhysRevLett.14.57,Senovilla:1998oua,Hawking:1970zqf} establishes that singularities are a generic outcome of gravitational collapse under the classical equations of General Relativity. These regions, where spacetime curvature becomes unbounded, signal a breakdown in the deterministic structure of the theory and raise deep concerns about its completeness. To preserve predictability, the cosmic censorship conjecture \cite{Penrose:1969pc} suggests that such singularities must always be hidden behind event horizons, rendering them inaccessible to distant observers.
This perspective has motivated sustained efforts to construct regular black hole models that avoid the formation of singularities altogether. Such models typically modify the matter content or the gravitational dynamics in the high-curvature regime, allowing for curvature invariants to remain finite throughout the spacetime. As a result, the study of non-singular black holes remains an important direction in gravitational research, offering potential insights into quantum gravity and the resolution of fundamental inconsistencies within classical theory.

The first known example of a regular black hole was proposed by Bardeen \cite{1968qtr..conf...87B}, who replaced the central singularity with a core supported by a specific form of matter, interpreted later as a magnetic charge arising from nonlinear electrodynamics. Instead of a singular region, the interior contained a smooth matter distribution resembling an anti de Sitter core. Since this initial proposal, many regular black hole solutions have been constructed using various matter contents \cite{PhysRevLett.96.031103,PhysRevLett.80.5056,Bronnikov:2005gm}, each aiming to preserve the essential features of a black hole while avoiding curvature divergence. In parallel, similar solutions have been developed within alternative theories of gravity \cite{Bueno:2024dgm,Junior:2023ixh,Carballo-Rubio:2021wjq,Berej:2006cc,Junior:2023qaq,Mazza:2023iwv}, where the gravitational dynamics is modified at short distances. Further efforts include the use of quantum-inspired corrections \cite{PhysRevD.76.104030,Modesto:2010rv,Fernandes:2023vux,Nicolini:2019irw,PhysRevD.106.L021901,PhysRevD.62.043008,Bonanno:2023rzk}, which attempt to capture effects of quantum gravity in an effective way, and approaches based on regularizing curvature invariants directly\cite{Carballo-Rubio:2019fnb}.

The concept of asymptotic safety has also received significant attention in recent years \cite{PhysRevD.62.043008,Bonanno:2023rzk,Eichhorn:2012va,PhysRevD.109.024045,Borissova:2022mgd,Pawlowski:2023dda}. Within this framework, gravity is expected to remain well defined at all energy scales due to the presence of a nontrivial ultraviolet fixed point. Building on earlier ideas by Markov and Mukhanov, Bonanno and collaborators \cite{Bonanno:2023rzk} recently introduced a new model of a regular black hole grounded in this approach. Their construction incorporates the antiscreening behavior of gravitational interactions at energy scales beyond the Planck threshold. By applying these principles, they derived an explicit spacetime geometry that describes the exterior region of a collapsing dust cloud, consistent with the expectations of asymptotically safe gravity.

Quantum corrections that remove the singularity inside a black hole are generally expected to act at Planck scale distances and are often considered too small to influence astrophysical observations. This belief stems from the fact that, far from the event horizon, the geometry tends to match the classical Kerr solution. However, in the region close to the horizon, even small departures from the Kerr geometry -- introduced by quantum effects or regular rotating black hole models -- can affect important physical quantities. Notably, jet power and radiative efficiency are determined by the structure of the spacetime near the black hole and depend on the position of the innermost stable circular orbit and the mechanism of energy extraction. As a result, precise measurements of these quantities in active galactic nuclei or black hole binaries may offer a possible way to detect subtle effects of quantum gravity.

Relativistic jets are a prominent feature observed in both active galactic nuclei and black hole X-ray binaries, such as microquasars. In the context of black hole X-ray binaries, two distinct types of jets are typically observed \cite{Fender:2004gg}. Steady jets are present during the hard state and can persist across a broad range of the source's luminosity. These jets are generally characterized by moderate relativistic speeds. In contrast, transient jets are seen in low-mass black hole X-ray binaries when the system transitions from the hard state to the soft state, with luminosities approaching the Eddington limit. These transient jets are often observed as high-velocity plasma blobs moving outward, with their origins believed to be near the event horizon of the black hole. As suggested by \cite{Narayan:2011eb}, the transient jets may be driven by the rotational energy of the black hole itself. Furthermore, since these jets appear at a well-defined luminosity, they hold potential as standard candles for distance measurement in astrophysical contexts.

The paper is organized as follows: In \ref{SEC:ASGI}, we have briefly discussed the theory of Asymptotic Safe Gravity and its influence on a regular rotating black hole . In \ref{SEC:3} we discuss about  the change in Radiative efficiency and the Relativistic JET power of a source in the context of quantum corrected spacetimes. In \ref{SEC:4} we discuss the constraint in $(\xi-a)$ parameter space with known observed information for six black holes. Finally in \ref{SEC:CONC}, we have discussed about our analysis and important inferences.

\section{Asymptotic Safe Gravity}\label{SEC:ASGI}

Asymptotic Safe Gravity (ASG) has emerged as one of the most compelling approaches to quantum gravity, wherein the gravitational interaction remains predictive and renormalizable at all energy scales due to the existence of a non-Gaussian ultraviolet (UV) fixed point. Unlike other approaches that require the introduction of action of new degrees of freedom, ASG preserves general covariance and classical field content, while implementing quantum corrections through scale-dependent couplings derived from renormalization group (RG) flow. In the deep UV regime, Newton's constant effectively diminishes with increasing energy scale, softening curvature singularities and enabling the construction of regular black hole geometries.

Bonanno et al.\cite{Bonanno:2023rzk}  extended the conceptual framework originally introduced by Markov and Mukhanov, proposing a mechanism in which the gravitational collapse of a dust cloud leads to a non-singular black hole configuration. The effective action incorporating scale-dependent couplings can be expressed as
\begin{flalign}\label{Eq:Action_ASG}
	\mathcal{S} = \frac{1}{2\kappa^2} \int \left[ R + 2\chi(\epsilon)\mathcal{L} \right] \sqrt{-g} \, \dd^4x,
\end{flalign}
where $\kappa^2 = 8\pi G_N$, $\chi(\epsilon)$ is an energy-scale-dependent coupling function, and $\mathcal{L}$ represents the matter Lagrangian. The scale $\epsilon$ denotes the relevant energy scale in the RG-improved spacetime, and the function $\chi(\epsilon)$ governs the way quantum corrections enter the gravitational sector.

Variation of this action with respect to the metric yields a modified set of Einstein equations:
\begin{flalign}
	R_{\mu\nu} - \frac{1}{2}g_{\mu\nu}R = 8\pi G(\epsilon)T_{\mu\nu} - \Lambda(\epsilon)g_{\mu\nu} \equiv T_{\mu\nu}^{\rm eff},
\end{flalign}
where $T_{\mu\nu}$ is the classical energy-momentum tensor for an ideal fluid:
\[
T_{\mu\nu} = \left\{ \epsilon + p(\epsilon) \right\} u_\mu u_\nu + p(\epsilon)g_{\mu\nu}.
\]
The energy dependence of the effective gravitational and cosmological couplings is encoded in:
\begin{subequations}
	\begin{flalign}
		8\pi G(\epsilon) &= \frac{d}{d\epsilon} \left( \chi(\epsilon) \epsilon \right), \\
		\Lambda(\epsilon) &= -\frac{d\chi}{d\epsilon} \epsilon^2.
	\end{flalign}
\end{subequations}

The renormalization group-improved Newtonian coupling, motivated by fixed-point behavior near the UV limit, takes the phenomenological form:
\begin{flalign}
	G(\epsilon) = \frac{G_N}{1 + \xi \epsilon},
\end{flalign}
where $\xi$ is a scale parameter governing the strength of quantum corrections. Its value is model-dependent and subject to constraint from theoretical consistency and observational data.

In the limit of gravitational collapse of a pressureless dust cloud ($p = 0$), the exterior spacetime metric adopts a static, spherically symmetric form:
\begin{flalign}\label{Eq:Metric}
	ds^2 = -F(r) dt^2 + \frac{dr^2}{F(r)} + r^2 \left( d\theta^2 + \sin^2\theta\, d\phi^2 \right),
\end{flalign}
with a lapse function given by:
\begin{flalign}
	F(r) = 1 - \frac{r^2}{3\xi} \ln\left( 1 + \frac{6M\xi}{r^3} \right),
\end{flalign}
where $M$ is the ADM mass of the black hole. This solution reduces to the Schwarzschild metric in the classical limit $\xi \rightarrow 0$. {\bf  In Ref.~\cite{Bonanno:2023rzk} the metric given in \ref{Eq:Metric} arises as the exterior part of a global spacetime constructed by matching a collapsing homogeneous dust interior to a static exterior geometry. The mass function is continuous across the boundary surface and takes the form  
\begin{equation}
	m(r) = \frac{r^{3}}{6 \xi} \ln\!\left(1+\frac{6 M\xi}{r^{3}}\right).
\end{equation}
If this exterior metric is taken in isolation and formally extended down to $r=0$, the Kretschmann scalar diverges and the geometry is not regular at the origin. However, this is not the physical situation described in Ref.~\cite{Bonanno:2023rzk}. The exterior solution has domain $r \geq r_{b}(t)=\tilde{r}_{b}a(t) > 0$, where $a(t)$ is the scale factor of the interior dust cloud. The central region $r<r_{b}(t)$ is filled with the homogeneous dust interior. In this region the Kretschmann scalar is proportional to $1/a^{n}(t)$ for some positive integer $n$, and since $a(t)$ never reaches zero at any finite comoving time, curvature invariants remain finite.  

Therefore, the complete spacetime obtained by joining the interior dust region with the exterior solution is free from curvature singularities. The designation of the solution as ``regular'' refers to this global construction, where the singularity at $r=0$ is never realized within the physical domain of the exterior geometry.}

\begin{figure}
	\centering
	\includegraphics[scale=0.7]{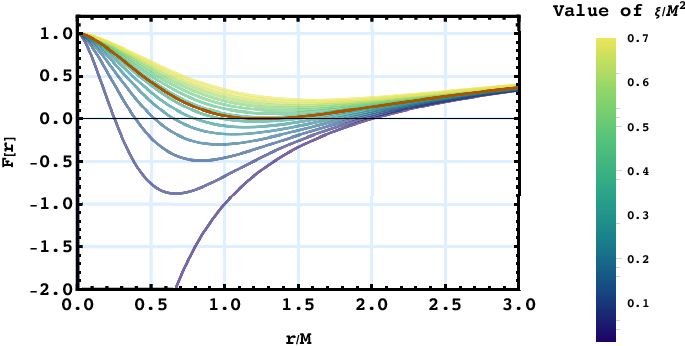}
	\caption{Variation of the lapse function $F(r)$ with radial coordinate $r$ for various values of the quantum correction parameter $\xi$. For $\xi = 0$, the Schwarzschild solution is recovered. Two-horizon behavior emerges for $\xi \lesssim \xi_{\rm crit}$.}\label{Fig:Lapse}. 
\end{figure}

An extremal configuration, characterized by coincident inner and outer horizons, arises for critical values of $\xi$
\begin{subequations}\label{Eq:Crit}
	\begin{flalign}
		\xi_{\rm crit} &= \frac{2}{3} y^2(3 + 2y)M^2, \\
		r_H^{\rm ext} &= -2yM, \\
		y &= W_0\left(-\frac{3}{2e^{3/2}}\right),
	\end{flalign}
\end{subequations}
where $W_0$ denotes the principal branch of the Lambert W function. Numerically, these yield $\xi_{\rm crit}/M^{2} \approx 0.4565 $ and $r_H^{\rm ext}/M \approx 1.2516 $. In \ref{Fig:Lapse}, the dependence of lapse function on the deformation parameter $\xi$ has been shown to confirm the critical value of the parameter.It is interesting to observed that depending upon the parameter value, the black hole can have two horizons.

\subsection{Regular Rotating Black Holes in ASG}

To account for rotation, the regular black hole solution is extended via a modified Kerr-like metric that incorporates ASG corrections. The resulting space-time geometry is given by (see \ref{App})
\begin{flalign}\label{Eq:Metric_kerr}
	ds^2 &= -\left[1 - \frac{2f(r)}{\Sigma(r,\theta)}\right] dt^2 + \frac{\Sigma(r,\theta)}{\Delta(r)} dr^2 + \Sigma(r,\theta) d\theta^2 \nonumber \\
	&\quad - \frac{4af(r)}{\Sigma(r,\theta)} \sin^2\theta\, dt\, d\phi + \frac{\mathcal{A}(r,\theta)}{\Sigma(r,\theta)}\sin^{2}\theta d\phi^2,
\end{flalign}
where the auxiliary functions are
\begin{subequations}
	\begin{flalign}
		\Sigma(r,\theta) &= r^2 + a^2 \cos^2\theta, \\
		f(r) &= \frac{1}{2} r^2 (1 - F(r)), \\
		\Delta(r) &= r^2 + a^2 - 2f(r), \\
		\mathcal{A}(r,\theta) &= (r^2 + a^2)^2 - a^2 \Delta(r) \sin^2\theta.
	\end{flalign}
\end{subequations}

Restricting to equatorial motion by setting $\theta = \pi/2$ simplifies the line element to:
\begin{flalign}
	ds^2_{\rm Eq} = &-F(r)\,dt^2 + \frac{r^2}{r^2 F(r) + a^2} dr^2 - 2a(1 - F(r)) dt\, d\phi\nonumber\\
	 &+ [r^2 + a^2 + a^2(1 - F(r))] d\phi^2.
\end{flalign}

\begin{figure}[h!]
	\centering
	\includegraphics[width=6cm]{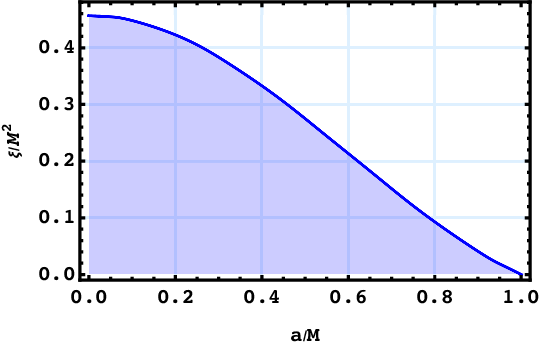}
	\caption{Parameter boundaries for horizon formation in the $(a, \xi)$ parameter space.}\label{Fig:PSpace}
\end{figure}

For further insights, the metric \eqref{Eq:Metric_kerr} can be recast in generalized Kerr-Schild coordinates
\begin{flalign}\label{Eq:GKS}
	ds^2 &= -\left[1 - \frac{2f(r)}{\Sigma}\right] dt'^2 + \left[1 + \frac{2f(r)}{\Sigma}\right] dr^2 + \frac{4f(r)}{\Sigma} dt'\, dr  \nonumber \\
	&\quad + \Sigma\, d\theta^2+ \left[\frac{(r^2 + a^2)^2 - a^2 \Delta \sin^2\theta}{\Sigma}\right] \sin^2\theta\, d\phi'^2 \nonumber \\
	&\hspace{0.5cm} - 2\left[1 + \frac{2f(r)}{\Sigma}\right] a \sin^2\theta\, dr\, d\phi' - \frac{4f(r)}{\Sigma} a \sin^2\theta\, dt'\, d\phi'.
\end{flalign}

In \ref{Fig:PSpace}, the parameter space has been shown for 
valid rotating black hole solution.

\section{Fundamental Framework: Radiative Efficiency and Relativistic Jet Power in Quantum-Corrected Rotating Spacetimes}\label{SEC:3}

The interplay between accretion dynamics and relativistic magnetohydrodynamics near black holes provides a powerful window into the structure of spacetime, particularly in the strong-field regime where quantum gravitational corrections may become non-negligible. In this section, we construct the theoretical foundation linking two complementary observables -- radiative efficiency of accretion disks and the power output of relativistic jets -- to the geometric and dynamical properties of rotating spacetimes modified by Asymptotic Safe Gravity (ASG). In this framework, Newton’s constant becomes scale-dependent, and quantum corrections are introduced through a deformation parameter $\xi$, altering the near-horizon geometry and affecting both geodesic motion and horizon-scale dynamics. We aim to understand how these corrections modify energy extraction processes and whether such modifications are observationally distinguishable from classical general relativity.

\subsection{Energy Extraction from Accretion: Modified Disk Efficiency in ASG}

The Novikov-Thorne model\cite{Novikov:1973kta} remains a reliable framework for studying geometrically thin, optically thick accretion disks in a broad class of stationary axisymmetric spacetimes. In the ASG context, the spacetime metric incorporates scale-dependent couplings, leading to corrections in the locations and energetics of circular orbits. For particles on equatorial circular geodesics, the four-velocity normalization condition,
\begin{equation}
	u^\mu u_\mu = -1,
\end{equation}
yields an effective potential formulation, where the quantum corrections modify the metric functions \(g_{\mu\nu}(r, \theta; \xi)\). The effective potential governing orbital motion takes the form\cite{Bambi:2017khi}
\begin{equation}
	V_{\text{eff}} = \frac{E^2 g_{\phi\phi} + 2 E L g_{t\phi} + L^2 g_{tt}}{g_{t\phi}^2 - g_{tt} g_{\phi\phi}} - 1,
\end{equation}
with \(E\) and \(L\) representing the specific energy and specific angular momentum. These conserved quantities depend on the angular velocity \(\Omega\) of the particle, given by:
\begin{align}
	E &= \frac{-g_{tt} - \Omega g_{t\phi}}{\sqrt{-g_{tt} - 2\Omega g_{t\phi} - \Omega^2 g_{\phi\phi}}}, \\
	L &= \frac{\Omega g_{\phi\phi} + g_{t\phi}}{\sqrt{-g_{tt} - 2\Omega g_{t\phi} - \Omega^2 g_{\phi\phi}}},
\end{align}
where the corrected angular velocity satisfies
\begin{equation}
	\Omega = \frac{-g_{t\phi,r} \pm \sqrt{(g_{t\phi,r})^2 - g_{\phi\phi,r} g_{tt,r}}}{g_{\phi\phi,r}}.
\end{equation}

The ISCO location \(r_{\text{ISCO}}\), a key radius where the disk truncates, is obtained by solving
\begin{equation}
	V_{\text{eff}} = 0, \quad \frac{dV_{\text{eff}}}{dr} = 0, \quad \frac{d^2V_{\text{eff}}}{dr^2} = 0.
\end{equation}
Once \(r_{\text{ISCO}}\) is determined, the radiative efficiency follows as:
\begin{equation}
	\eta_{NT} \equiv 1 - E_{\text{ISCO}}(a,\xi).
\end{equation}

\begin{figure}[h!]
	\centering
	\subfloat[\emph{Variation of ISCO radius with change in spin parameter for full range of parameter space.}]{{	\includegraphics[width=8cm]{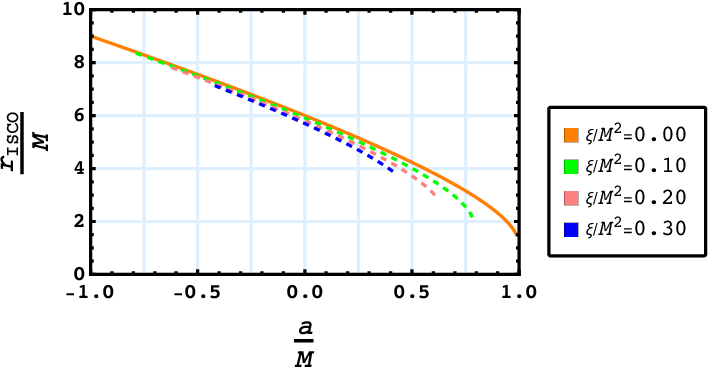}}\label{Fig:Pro_a}}
	\qquad
	\subfloat[\emph{Variation of ISCO radius with change in spin parameter for narrow range of parameter space.}]{{\includegraphics[width=8cm]{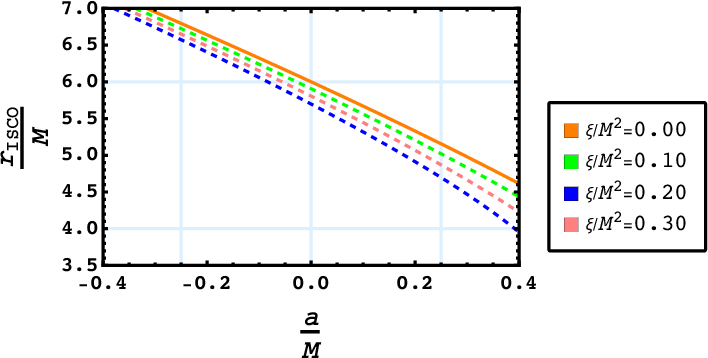}}\label{Fig:Pro_b}}
	\qquad
	\subfloat[\emph{Variation of ISCO radius with change in deformation parameter.}]{{\includegraphics[width=8cm]{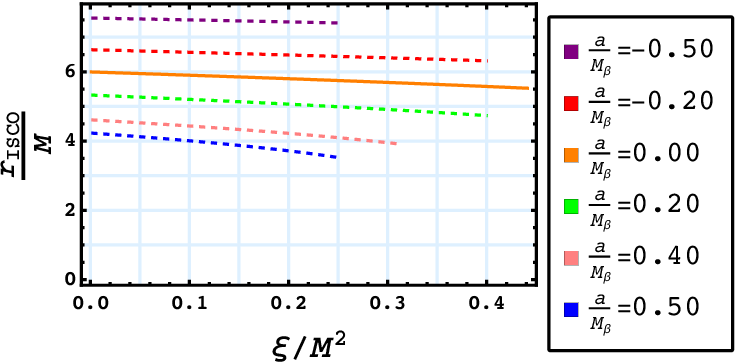}}\label{Fig:Pro_c}}
	\caption{Variation of disk radiative efficiency $\eta$ with spin and ASG parameter $\xi$ for two representative cases.}
	\label{Fig:ISCOa}
\end{figure}

\begin{figure}[h!]
	\subfloat[\emph{Variation of ISCO radius with change in spin parameter $a$ and deformatioon parameter $\xi$ for full range of parameter space.}]{{\includegraphics[width=8cm]{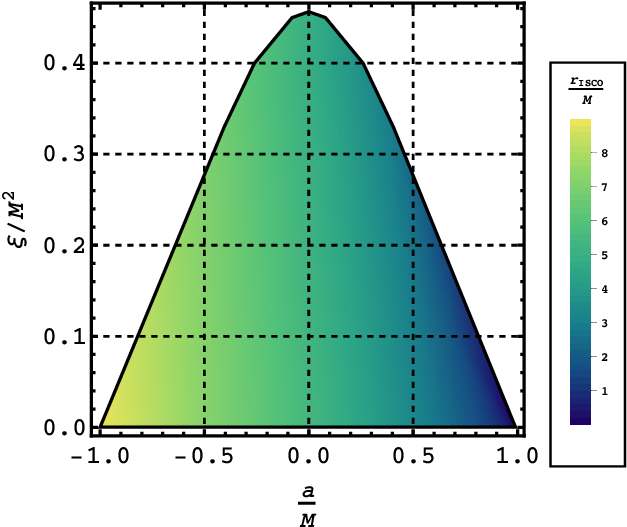}}}
	\caption{Variation of disk radiative efficiency $\eta$ with spin and ASG parameter $\xi$ for two representative cases.}
	\label{Fig:ISCO}
\end{figure}

\begin{figure}[htbp!]
	\centering
	\subfloat[\emph{\bf Radiative Efficiency shown in full range of the parameter space}]{{\includegraphics[width=7.5cm]{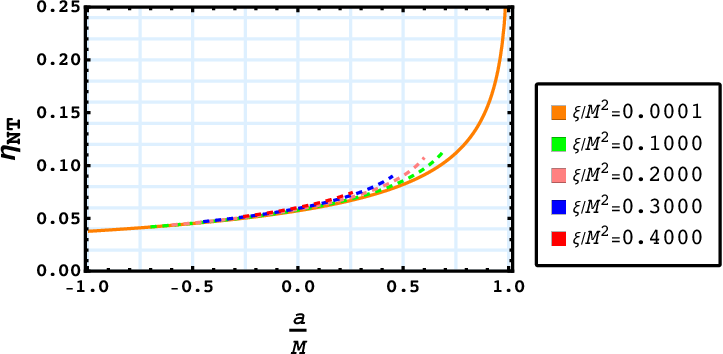}}}
	\qquad
	\subfloat[\emph{\bf Radiative Efficiency has been shown in a narrow region}]{{\includegraphics[width=7.5cm]{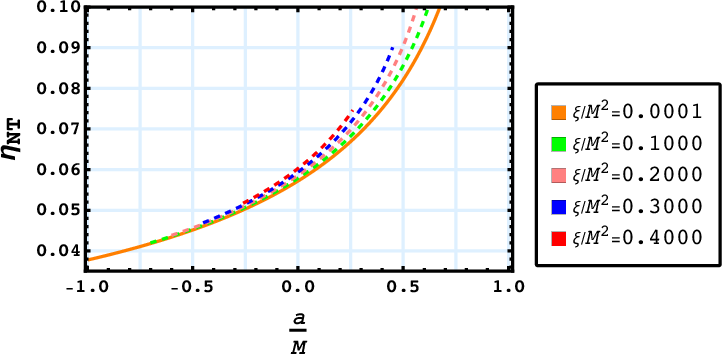}}}
	\caption{{\bf Variation of the disk radiative efficiency with the spin parameter $a$ for the same physical case. 
			The upper panel shows the full parameter range, while the lower panel zooms into a shorter range for clarity.}
	}
	\label{Fig:TSS_Variation}
\end{figure}

In ASG-modified spacetimes, the presence of a running Newton’s constant generically shifts the ISCO radius outward for positive values of the deformation parameter \(\xi\), thereby reducing the efficiency \(\eta_{NT}\) at a given spin. This behavior reflects the impact of quantum gravity corrections on geodesic structure and enables radiative measurements to indirectly probe deviations from the classical Kerr solution. The ISCO radius associated with the rotating black hole in ASG is given in \ref{Fig:ISCO}. {\bf The explicit dependence of the ISCO radius on the spin parameter $a$ and the deformation parameter $\xi$ is illustrated in \ref{Fig:ISCOa}. From \ref{Fig:Pro_a} and \ref{Fig:Pro_b}, it can be observed that for a fixed value of $\xi$, the ISCO radius is larger for retrograde orbits compared to prograde orbits. In \ref{Fig:Pro_c}, the variation of the ISCO radius with respect to the deformation parameter is presented for fixed values of the spin parameter $a$. This indicates that for a given spin, the rate of change of the ISCO radius with respect to the deformation parameter is very small. Nevertheless, this effect cannot be ignored, since the radiative efficiency shows significant variation with changes in the deformation parameter for fixed spin. The dependence of the radiative efficiency on the spin parameter, for different values of the deformation parameter, is shown in \ref{Fig:TSS_Variation}. Within the considered parameter range, one can notice that for a fixed spin of the black hole, the radiative efficiency increases with increasing values of the deformation parameter. However, the maximum radiative efficiency is achieved in the case of the Kerr black hole.
}

\subsection{Relativistic Jet Power in the Presence of Running Couplings}

Complementing disk emission, relativistic jets provide an independent observational window into the near-horizon region, where quantum-corrected gravity may significantly affect energy extraction mechanisms. The Blandford-Znajek (BZ) process remains the leading theoretical framework for jet production via magnetic extraction of rotational energy. In ASG-corrected spacetimes, the force-free magnetosphere structure is retained, but key quantities such as the horizon’s angular velocity are modified by quantum effects.

The electromagnetic stress-energy tensor in a force-free environment is given by
\begin{equation}
	T^{\mu\nu}_{\text{EM}} = F^{\mu\alpha} F^\nu{}_\alpha - \frac{1}{4} g^{\mu\nu} F_{\alpha\beta} F^{\alpha\beta},
\end{equation}
where \(F_{\mu\nu} = \partial_\mu A_\nu - \partial_\nu A_\mu\) is the field strength tensor. For stationary, axisymmetric field configurations with rigid rotation of field lines, the angular velocity \(\omega(r, \theta)\) satisfies
\begin{equation}
	\frac{\partial_r A_t}{\partial_r A_\phi} = \frac{\partial_\theta A_t}{\partial_\theta A_\phi} = -\omega(r, \theta).
\end{equation}

The total energy flux carried by the jet is extracted at the event horizon, with the Poynting flux \(T^r_t\) contributing to the net power\cite{1977MNRAS.179..433B}
\begin{equation}
	P_{\text{BZ}} = 4\pi \int_0^{\pi/2} \sqrt{-g} \, T^r_t \, d\theta,
\end{equation}
where
\begin{equation}
	T^r_t = 2 r_H M \sin^2\theta (B_r)^2 \omega(\Omega_H - \omega)\big|_{r=r_H}.
\end{equation}

The angular velocity of the horizon in ASG depends on both the spin and quantum deformation parameter:
\begin{equation}
	\Omega_H = -\left. \frac{g_{t\phi}}{g_{\phi\phi}} \right|_{r=r_H(a, \xi)}.
\end{equation}
The explicit form of \(\Omega_H\) varies with the ASG model employed, but the general trend is a suppression of angular velocity with increasing \(\epsilon\), due to metric softening near the horizon.

Jet power thus scales as \cite{2010ApJ...711...50T}
\begin{equation}
	P_{\text{BZ}} = k \Phi_{\text{tot}}^2 \, \Omega_H^2(a, \xi),
\end{equation}
where $k$ is $1/6\pi$ if split monopole profile is considered and $k=0.044$ is considered if paraboloidal profile is taken. The logarithmic form is given by
\begin{equation}
	\log P_{\text{BZ}} = \log K + 2 \log \Omega_H(a, \xi),
\end{equation}
with \(K = k \Phi_{\text{tot}}^2\), facilitates direct comparison with observational data.

In summary, the ASG-induced quantum corrections simultaneously affect both the efficiency of radiative emission from the accretion disk and the power output of relativistic jets by altering the ISCO structure and the angular velocity at the event horizon. The two observables -- \(\eta_{NT}(a, \xi)\) and \(P_{\text{BZ}}(a, \xi)\) -- thus act as independent, complementary probes of quantum gravity near rotating black holes. In the following sections, we use these quantities to place constraints on the ASG parameter space using data from stellar-mass black hole binaries.

\section{Constraints from Observational Data in Asymptotically Safe Rotating Black Holes}
\label{SEC:4}

Asymptotic safety offers a promising ultraviolet completion of gravity, wherein the gravitational couplings become scale-dependent and flow toward a non-Gaussian fixed point at high energies. Black hole spacetimes within this framework acquire quantum corrections that can subtly alter the geodesic structure, the horizon properties, and the spacetime curvature at small scales. In this section, we aim to constrain the parameters of a rotating black hole in asymptotically safe gravity (ASG) using two observationally accessible quantities: the radiative efficiency of the accretion disk and the power of relativistic jets. We investigate whether a unified region in the parameter space of spin and quantum deformation (encoded in the running of Newton’s constant or additional ASG-specific charges) can reproduce both observed quantities for several stellar-mass black hole binaries.

The radiative efficiency of an accretion disk encodes the gravitational binding energy lost by matter as it spirals from infinity down to the innermost stable circular orbit (ISCO). It is calculated from the specific energy \(E_{\text{ISCO}}\) along equatorial circular geodesics
\begin{equation}
	\eta_{NT} = 1 - E_{\text{ISCO}}. \label{etais_asg}
\end{equation}
In ASG-corrected spacetimes, both the location and energy at the ISCO depend on the black hole spin \(a\) and on the quantum correction parameter, typically denoted  \(\xi\), characterizing the running of Newton’s constant \(G(\epsilon)\). \ref{Table1} lists the black hole mass \(M\), distance \(D\), inclination angle \(i\), spin \(a\), and the Kerr-assumed estimate of \(\eta\) for six well-studied X-ray binary systems. These values serve as benchmarks to assess whether the quantum-corrected rotating black hole metrics can still fit the data when the spacetime deviates from classical general relativity.

{%\onecolumn
%\begin{widetext}

\begin{table}[ht]
	\centering
	\caption{Parameters of the black hole binaries analyzed in the context of ASG. The radiative efficiency \(\eta\) is derived using ~\eqref{etais_asg} under the classical Kerr assumption. Quantum corrections may shift the effective value of \(\eta\) for a given spin.}
	\label{Table1}
	\resizebox{0.5\textwidth}{!}{
	\begin{tabular}{|c|c|c|c|c|c|}
		\hline
		BH Source & $M \, (M_\odot)$ & $D \, (\text{kpc})$ & $i^\circ$ & $a$ & $\eta$ \\
		\hline
		A0620-00 & $6.61 \pm 0.25$ & $1.06 \pm 0.12$ & $51.0 \pm 0.9$ & $0.12 \pm 0.19$ ~\cite{Gou_2010}& $0.061^{+0.009}_{-0.007}$ \\
		H1743-322 & $8.0$ & $8.5 \pm 0.8$ & $75.0 \pm 3.0$ & $0.2 \pm 0.3$ ~\cite{2012ApJ...745L...7S}& $0.065^{+0.017}_{-0.011}$ \\
		XTE J1550-564 & $9.10 \pm 0.61$ & $4.38 \pm 0.5$ & $74.7 \pm 3.8$ & $0.34 \pm 0.24$~\cite{2012ApJ...745L...7S}& $0.072^{+0.017}_{-0.011}$ \\
		GRS 1124-683 & $11.0^{+2.1}_{-1.4}$ & $4.95^{+0.69}_{-0.65}$ & $43.2^{+2.1}_{-2.7}$ & $0.63^{+0.16}_{-0.19}$~\cite{Chen_2016}& 
		$0.095^{+0.025}_{-0.017}$ \\
		GRO J1655-40 & $6.30 \pm 0.27$ & $3.2 \pm 0.5$ & $70.2 \pm 1.9$ & $0.7 \pm 0.1$~\cite{Shafee_2005}& $0.104^{+0.018}_{-0.013}$ \\
		GRS 1915+105 & $12.4^{+1.7}_{-1.9}$ & $8.6^{+2.0}_{-1.6}$ & $60.0 \pm 5.0$ & $a_{*} > 0.98$~\cite{McClintock_2006}& $\eta > 0.234$ \\
		\hline
	\end{tabular} }  
\end{table}
%\end{widetext}
}
%\twocolumn

Alongside disk energetics, the presence of relativistic jets offers a second diagnostic that directly connects to the spacetime structure near the horizon. Following standard assumptions, the jet is modeled as a pair of symmetric outflows (plasmoids) traveling with bulk Lorentz factor \(\Gamma\). The relation between the observed and intrinsic radio flux is governed by the Doppler factor \(\delta\), as:
\begin{align}
	\frac{S_\nu}{S_{\nu,0}} &= \delta^{3 - \alpha}, \label{boost_asg} \\
	\delta &= \left[\Gamma (1 - \beta \cos i)\right]^{-1}, \label{doppler_asg}
\end{align}
where \(\alpha\) is the spectral index, \(\beta = v/c\), and \(i\) is the jet inclination. For sources with modest Lorentz factors and intermediate inclinations, this correction can significantly affect the inferred jet power.

The observed (proxy) jet power is computed using the Doppler-corrected peak radio flux at 5 GHz:
\begin{equation}
	P_{\text{jet}} = \left(\frac{\nu}{5~\text{GHz}}\right) \left(\frac{S_{\nu,0}^{\text{tot}}}{\text{Jy}}\right) \left(\frac{D}{\text{kpc}}\right)^2 \left(\frac{M}{M_\odot}\right)^{-1}. \label{Pjet_asg}
\end{equation}

\begin{table}[ht]
	\centering
	\caption{Jet power estimates (proxy units: kpc$^2$~GHz~Jy~$M_\odot^{-1}$) for two values of Lorentz factor \(\Gamma\). These are derived from observed radio peaks and corrected for relativistic beaming.}
	\label{Table2}
	\begin{tabular}{|c|c|c|c|}
		\hline
		BH Source & \((S_{\nu,0})^{5\,\text{GHz}}_{\text{max}}\) (Jy) & \(P_{\text{jet}}|_{\Gamma=2}\) & \(P_{\text{jet}}|_{\Gamma=5}\) \\
		\hline
		A0620-00 & 0.203 & 0.13 & 1.6 \\
		H1743-322 & 0.0346 & 7.0 & 140 \\
		XTE J1550-564 & 0.265 & 11 & 180 \\
		GRS 1124-683 & 0.45 & 3.9 & 380 \\
		GRO J1655-40 & 2.42 & 70 & 1600 \\
		GRS 1915+105 & 0.912 & 42 & 660 \\
		\hline
	\end{tabular}
\end{table}

In the framework of Asymptotic Safe Gravity, the horizon angular velocity \(\Omega_H\) receives scale-dependent corrections that alter the jet-launching efficiency. Under the assumption that the Blandford-Znajek relation remains valid with effective couplings, the theoretical jet power is modeled as:
\begin{equation}
	\log P = \log K + 2 \log \Omega_H, \label{Pfit_asg}
\end{equation}
where \(K = k \Phi^2_{\text{tot}}\) is assumed universal across sources, following empirical fits. Values of \(\log K = 2.94 \pm 0.22\) (\(\Gamma = 2\)) and \(\log K = 4.19 \pm 0.22\) (\(\Gamma = 5\)) are adopted from Ref.~\cite{Middleton:2014cha}. The spacetime dependence enters through \(\Omega_H(a, \epsilon)\), where \(\epsilon\) denotes the quantum deformation parameter in ASG. The horizon angular velocity is computed from the ASG-corrected metric, which typically modifies the location of the event horizon and the frame-dragging coefficients near it.

By comparing \ref{etais_asg} with the observational efficiencies and \ref{Pfit_asg} with the jet powers listed in ~\ref{Table2}, we numerically scan the \((a, \xi)\) parameter space to identify regions consistent with both observables. This enables a joint constraint on the spin and quantum deformation, with allowed bands defined by radiative and jet fits. The intersection of these bands constitutes a physically viable regime in which ASG-corrected rotating black holes can replicate the electromagnetic properties of real astrophysical systems.

\section{Joint Constraints from Radiative Efficiency and Jet Power}
\subsection{ Source A0620--00}
%\subsubsection{A0620-00}

%\subsubsection*{ in Asymptotic Safe Gravity: }

The black hole candidate A0620--00 offers a low-spin, low-luminosity system ideally suited for testing quantum-corrected gravity models. Within the Kerr framework, continuum-fitting methods (CFM) estimate the spin parameter of this system as $a = 0.12 \pm 0.19$ (68\% Confidence Level)~\cite{Gou_2010} translating into a Novikov-Thorne radiative efficiency of $\eta _{NT}= 0.061^{+0.009}_{-0.007}$ (see ~\ref{Table1}). In the framework of Asymptotic Safe Gravity (ASG), such a measurement can be reinterpreted as a constraint on the spin parameter $a$ and the ASG deformation parameter $\xi$, which governs the scale-dependent behavior of Newton’s constant. 

\begin{figure*}[htbp!]
	\centering
	\subfloat[]{\includegraphics[width=0.35\linewidth]{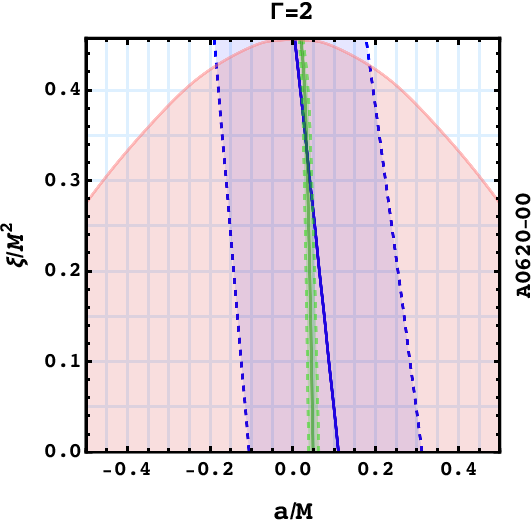}} 
	\qquad
	\subfloat[]{\includegraphics[width=0.35\linewidth]{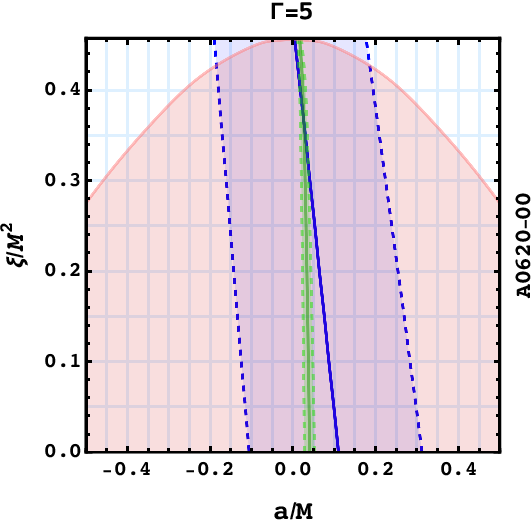}}
	\caption{\emph{Constraints on the spin parameter $a$ and the ASG deformation parameter $\xi$ for A0620--00. Blue shaded regions correspond to values consistent with the observed radiative efficiency $\eta_{NT}= 0.061^{+0.009}_{-0.007}$. Green shaded regions denote $(a, \xi)$ combinations reproducing the observed jet power within $0.3\, \text{dex}$ uncertainty. The overlap identifies parameter ranges compatible with both observables under Asymptotic Safe Gravity. The Red shaded region shows the allowed parameter space for a regular black hole in ASG.}}
	\label{fig:asg_constraints}
\end{figure*}

\ref{fig:asg_constraints} displays the allowed parameter space for A0620--00 under ASG, derived from simultaneous matching of the observed radiative efficiency and relativistic jet power. The left panel corresponds to a Lorentz factor $\Gamma = 2$ and the right to $\Gamma = 5$. In both panels, the blue shaded region represents the $(a, \xi)$ pairs that reproduce the observed efficiency $\eta_{NT} = 0.061$ within its error bounds. These curves serve as a quantum-deformed generalization of the CFM constraints under the Kerr hypothesis. As expected, when $\xi \to 0$, the central value of the blue region converges to the classical Kerr value $a \simeq 0.12$. For positive $\xi$, however, smaller spin values remain compatible due to the outward shift of the ISCO, which reduces $\eta_{NT}$  for a given $a$. 
Overlaying the blue region is the green-shaded band, which represents the set of $(a, \xi)$ configurations that reproduce the observed jet power $P_{\text{jet}}$ at the central value with an uncertainty of $0.3 \, \text{dex}$ (based on ~\ref{Table2}). For $\Gamma = 2$ (left), the central jet power is matched by both classical Kerr and ASG models, although the latter allows for slightly lower values of $a$ as $\xi$ increases. The green contour shifts towards the $a=0$ axis for $\Gamma = 5$ (right), indicating the decrease in required spin to maintain the same jet power under more relativistic beaming conditions.

A key observation is the intersection between the efficiency-based (blue) and jet-based (green) regions. This overlap identifies the physically viable configurations within ASG that satisfy both observational constraints. For A0620--00, the overlap is concentrated around $a \simeq 0.05 \pm 0.02$ and $0 < \xi \lesssim 0.5$. These values define a bounded region in the $(a, \xi)$ plane where quantum-corrected rotating black holes remain compatible with both the thermal and non-thermal emissions of the source. Notably, the classical Kerr metric with low spin lies within this region, but the ASG interpretation allows a continuous degeneracy with respect to the quantum parameter $\xi$.
Taken together, these results suggest that Asymptotic Safe Gravity can simultaneously account for the radiative and jet-related features of A0620--00, provided that the deformation parameter remains modest. The analysis highlights how low-spin systems can serve as sensitive probes for quantum gravity corrections in the strong-field regime.

\subsection{ Source H1743-322 : }

\begin{figure}[htbp!]
	\centering
	\subfloat[\emph{} ]{{\includegraphics[width=6cm]{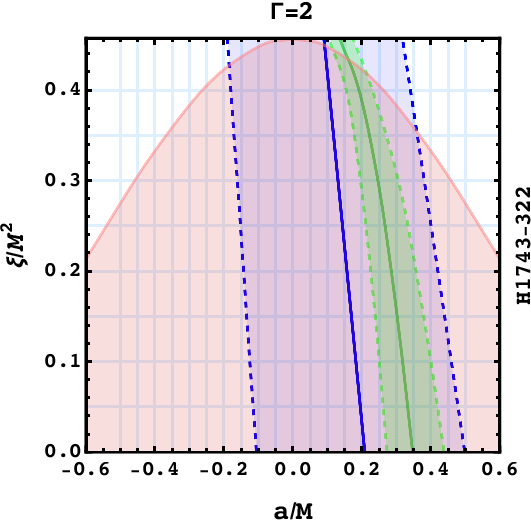}}\label{Fig:PE_2a}}
	\qquad
	\subfloat[\emph{}]{{\includegraphics[width=6cm]{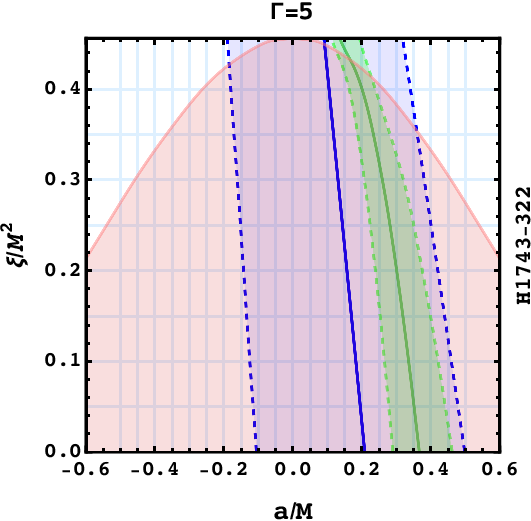}}\label{Fig:PE_2b}}
	\caption{\emph{Jet power and radiative efficiency constraints for the black hole system H1743-322 have been depicted. In each panel, the green shaded area represents the combinations of $\xi$ and spin $a$ that successfully account for the observed jet power, considering the uncertainties, for (a) Lorentz factor $\Gamma = 2$ and (b) $\Gamma = 5$. The blue shaded region corresponds to the parameter values that are consistent with the observed radiative efficiency $\eta_{NT}$, as reported in \ref{Table1}. A detailed explanation of the physical implications of these regions is provided in the main text.
	}}
	\label{Fig:Parameter_Estimation_2}
\end{figure}

{The spin of the black hole candidate in H1743-322 has been estimated using the continuum fitting method. The best-fit value corresponds to a spin parameter of approximately $a \approx 0.2$, with an uncertainty of $\pm 0.3$ at the 68 percent confidence level~\cite{Steiner_2011}. When extended to the 90 percent confidence level, the spin is constrained within the interval $-0.3 < a < 0.7$. Based on this range, the radiative efficiency is inferred to be $\eta_{NT} = 0.065^{+0.017}_{-0.011}$ at the 68 percent confidence level.
	To assess how these observational results relate to our model based on asymptotically safe gravity, we have analyzed the parameter space that reproduces the observed radiative efficiency of H1743-322. For the standard Kerr case, the central spin value of $a \approx 0.2$ is consistent with observational bounds. When quantum corrections from asymptotically safe gravity are incorporated, a shift in the allowed spin range is observed.

	The blue shaded region in ~\ref{Fig:Parameter_Estimation_2} represents the set of values in the $(\xi, a)$ parameter space that can explain the observed radiative efficiency within its reported uncertainties. The blue contours correspond to theoretical predictions obtained by matching the observed value of $\eta_{\text{NT}}$ with the model radiative efficiency as given in ~\ref{etais_asg}. The solid blue line represents the contour for the central value of $\eta_{\text{NT}}$, while the dotted blue curves indicate the boundaries set by the observational error margins, as documented in ~\ref{Table1}.
	From the figure, it becomes apparent that the entire range of the parameter $\xi$ under consideration is consistent with the observed radiative efficiency. The astrophysical object under study is also known to exhibit strong ballistic jets. The corresponding jet power values for Lorentz factor $\Gamma = 2$ and $\Gamma = 5$ are presented in ~\ref{Table1}, each associated with an uncertainty of approximately $0.3\,\rm dex$ relative to the central estimate.
	The green shaded area in ~\ref{Fig:Parameter_Estimation_2} denotes the allowed region in the $(\xi, a)$ space that accounts for the measured jet power, including both upper and lower bounds arising from the uncertainties. The definitions of the blue solid and dotted lines remain consistent with those described previously.
	~\ref{Fig:PE_2a} and~\ref{Fig:PE_2b} depict the constraints derived from jet power measurements for $\Gamma = 2$ and $\Gamma = 5$, respectively. From these panels, it is evident that the range $\xi \lesssim 0.42$ can simultaneously describe both the observed jet power and the radiative efficiency, thus indicating a region of compatibility. The intersection of the blue and green shaded regions highlights the values of $\xi$ and $a$ that are capable of jointly accounting for both observational signatures.

}

\subsection{Source XTE J1550-564}

\begin{figure}[htbp!]
	\centering
	\subfloat[\emph{} ]{{\includegraphics[width=6cm]{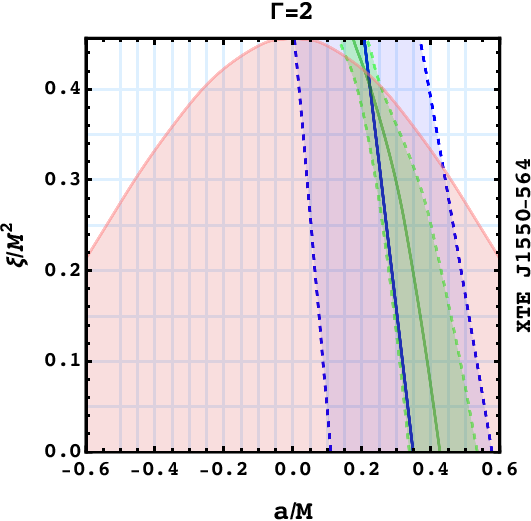}}\label{Fig:PE_3a}}
	\qquad
	\subfloat[\emph{}]{{\includegraphics[width=6cm]{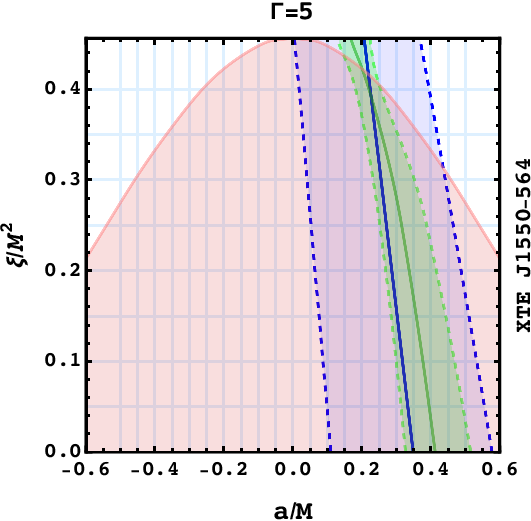}}\label{Fig:PE_3b}}
	\caption{\emph{Parameter constraints for the black hole binary XTE J1550$-$564 have been shown. The green regions in panels (a) and (b) show the combinations of $\xi$ and spin $a$ that yield theoretical jet power consistent with the observed radio luminosity, after accounting for Doppler boosting with Lorentz factors $\Gamma = 2$ and $\Gamma = 5$, respectively. The blue shaded area marks the parameter space where the radiative efficiency predicted by the model aligns with the measured value listed in ~\ref{Table1}. The green zone outlines the range of $\xi$ and $a$ that support the presence of a physical event horizon, ensuring the configuration corresponds to a bona fide black hole. Further insights and interpretations can be found in the main text.}}
	\label{Fig:Parameter_Estimation_3}
\end{figure}

{

	Among the well-studied microquasar systems, XTE~J1550$-$564 stands out as a low-mass X-ray binary harboring a stellar-mass black hole with a dynamically measured mass of $9.1 \pm 0.61\,M_\odot$~\cite{ steiner:2010bt}. The companion star in this system is classified as a late-G or early-K spectral type, typical of low-mass stellar donors undergoing Roche lobe overflow~\cite{Orosz:2002uh}. The orbital period of the binary has been determined to be approximately $1.55$ days, reflecting the tight dynamical coupling between the compact object and its donor star. The source lies at a distance of $4.38^{+0.58}_{-0.41}$ kpc from Earth and exhibits an orbital inclination of $74.7^\circ \pm 3.8^\circ$ relative to the line of sight~\cite{Orosz2011ANID}.
	
	The dimensionless spin parameter $\widetilde{a}=a/M$ of the black hole in this system has been constrained through both the Continuum Fitting (CF) method and the relativistic Fe-line profile modeling. The CF analysis yields a spin range of $-0.11 < a < 0.71$ at 90\% confidence, with a most probable value around $a = 0.34$~\cite{Steiner2010TheSO}. In contrast, the Fe-line method provides a slightly higher estimate, with $a = 0.55^{+0.15}_{-0.22}$~\cite{Steiner2010TheSO}. For consistency, we adopt the CF-based spin value in our analysis, as recorded in ~\ref{Table1}, and compute the corresponding radiative efficiency $\eta_{NT}$ from this estimate following the BZ-formalism in alternative gravity theories~\cite{Pei:2016kka}.
	Observations at radio wavelengths reveal a significant flux density of $0.265$ Jy at 5 GHz for this source. Assuming Lorentz factors $\Gamma = 2$ and $\Gamma = 5$, representative of mildly and highly relativistic jet outflows respectively, the associated kinetic jet powers have been calculated and listed in ~\ref{Table2}, with systematic uncertainties of approximately $0.3$ dex~\cite{Narayan:2011eb,Middleton:2014cha}. These observationally derived jet powers are juxtaposed with our model predictions in ~\ref{Fig:PE_3a} and~\ref{Fig:PE_3b}. The regions in the $\xi$--$a$ parameter space that are consistent with the observed jet power, accounting for errors, are shown as green-shaded zones in the respective figures.
	
	Simultaneously, the constraint from the observed radiative efficiency $\eta_{NT}$ is illustrated by the blue-shaded regions in both figures. Notably, the analysis reveals that values of $\xi/M^{2} > 0.44$ fail to reproduce the observed jet power, thereby placing an upper bound on this parameter. However, no such restriction arises from radiative efficiency constraints alone (with error bars). Interestingly, the combined observational bounds suggest that the maximal permissible value of $\xi/M^{2}$ is approximately $0.44$, a result that is consistent across both jet models.
	Furthermore, when $\xi= 0$, the range of spin parameters required to match the jet power aligns well with the spin interval derived from the CF method, thus reinforcing the internal consistency of the theoretical framework employed.

}

\subsection{Source GRS 1124 -683}

\begin{figure}[htbp!]
	\centering
	\subfloat[\emph{} ]{{\includegraphics[width=6cm]{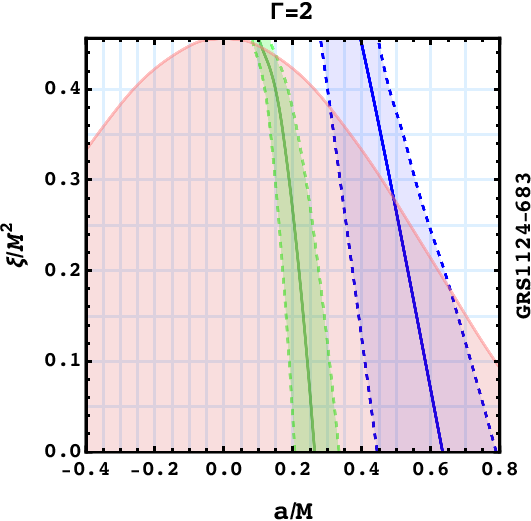}}\label{Fig:PE_4a}}
	\qquad
	\subfloat[\emph{}]{{\includegraphics[width=6cm]{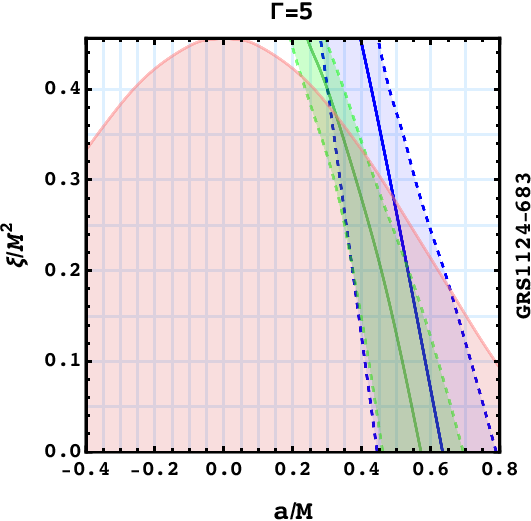}}\label{Fig:PE_4b}}
	\caption{\emph{Parameter space analysis for the black hole candidate RS 1124$-$683. The  'panel (a)' illustrates that when the jet Lorentz factor is taken as $\Gamma = 2$, no region  allows for a consistent explanation of both the observed jet power and radiative efficiency. In contrast, the  'panel (b)'-- corresponding to $\Gamma = 5$ -- reveals a region of overlap between the constraints from the two observables, indicating that such a configuration is compatible with the data. These findings suggest that high-speed jet ejection plays a crucial role in reconciling the source’s emission characteristics. For a detailed physical discussion, refer to the main text.
	}}\label{Fig:Parameter_Estimation_4}
\end{figure}

{
	
	The system under consideration is an X-ray binary comprising a stellar-mass black hole and a K-type main sequence companion star. The compact object has an estimated mass of $11.0^{+2.1}_{-1.4}\,M_\odot$~\cite{Wu:2015ioy}, while the orbital period of the binary is approximately $10.4$ hours~\cite{1992ApJ...399L.145R}, suggesting a tight configuration. The distance to the source has been measured as $D = 4.95^{+0.69}_{-0.65}$ kpc, and the inclination angle of the orbit is $i = 43.2^{+2.1}_{-2.7}$ degrees~\cite{Wu:2015ioy}. 
	Using the Continuum Fitting method applied to the thermal disk spectrum, the spin parameter $a$ of the black hole has been estimated to be $0.63^{+0.16}_{-0.19}$~\cite{Chen:2015mvc}. This spin value is used to compute the radiative efficiency $\eta_{NT}$ as listed in \ref{Table1}, following standard accretion disk theory. The corresponding region in the $\xi$--$a$ parameter space that is consistent with the observed $\eta_{NT}$ is shown as the blue shaded area in ~\ref{Fig:Parameter_Estimation_4}, indicating that the upper bound on $\widetilde{\xi}=\xi/M^{2}$ for radiative consistency is approximately $0.38$.
	
	Radio observations at $5$ GHz have been employed to infer the kinetic power of the jet for two assumed values of the Lorentz factor, $\Gamma = 2$ and $\Gamma = 5$. The inferred jet powers, along with an uncertainty of $0.3$ dex, are reported in ~\ref{Table2} \cite{Narayan:2011eb,Middleton:2014cha}. 
	A comparison of the jet power constraints, illustrated by the green regions in ~\ref{Fig:Parameter_Estimation_4}, reveals a clear distinction between the two assumed values of the Lorentz factor. In the left panel, which corresponds to $\Gamma = 2$, the spin parameter begins around $a \simeq 0.26^{+0.11}_{-0.08}$ for the standard Kerr geometry and decreases to approximately $a \simeq 0.12^{+0.04}_{-0.03}$ when the ASG parameter is set to $\xi/M^{2} = 0.44$. On the other hand, the right panel, associated with $\Gamma = 5$, starts from a significantly higher spin value of $a \simeq 0.57^{+0.17}_{-0.15}$ in the absence of deformation parameter  and drops gradually to $a \simeq 0.38^{+0.04}_{-0.08}$ for $\xi/M^{2} = 0.32$.
	When we consider both jet power and radiative efficiency as observational constraints, an important difference arises between the two panels. In the case of $\Gamma = 2$, no common overlap exists between the red and blue shaded regions, indicating that no combination of $\xi$ and $a$ can simultaneously explain both observables. However, for $\Gamma = 5$, the shaded regions intersect, highlighting a set of $(\xi, a)$ values that are consistent with both the measured jet power and the radiative efficiency.
	
	This result strongly suggests that the central object is more accurately modeled as  regular  rotating black  in ASG, and that the jet emission is likely to involve relativistic outflows with a Lorentz factor closer to $\Gamma = 5$. The presence of overlapping regions in this case supports the plausibility of a high-speed jet launching mechanism in the system under consideration.

}

\subsection{\bf Source: GRO J1655-40 }

The galactic microquasar GRO~J1655$-$40 is a well-studied X-ray binary system containing a black hole with a dynamically determined mass of $M = 6.3 \pm 0.5\,M_\odot$~\cite{Greene:2001wd}. Its companion star has been classified as an F-type main sequence object with a mass of $M_S = 2.34 \pm 0.12\,M_\odot$, and the orbital period of the system is measured to be $2.62$ days~\cite{Orosz:1996cg}. The distance to the source has been estimated as $D = 3.2 \pm 0.5$ kpc~\cite{1995Natur.375..464H}, and the orbital inclination is found to be $i = 70.2^\circ \pm 1.9^\circ$~\cite{Greene:2001wd}.
Despite the extensive observational data, the spin of the black hole in GRO~J1655$-$40 remains a matter of ongoing debate. The Continuum Fitting approach suggests a moderate spin in the range $a \sim 0.65$ to $0.75$~\cite{Shafee:2005ef}, whereas estimates based on the relativistically broadened iron line indicate a much higher value, $a > 0.9$~\cite{10.1111/j.1365-2966.2009.14622.x}. In addition, constraints derived from the analysis of quasi-periodic oscillations in the X-ray power spectrum yield significantly different estimates, with the black hole mass and spin inferred as $M = 5.31 \pm 0.07\,M_\odot$ and $a = 0.290 \pm 0.003$ respectively~\cite{Motta:2013wga}. 

\begin{figure}[htbp!]
	\centering
	\subfloat[\emph{} ]{{\includegraphics[width=6cm]{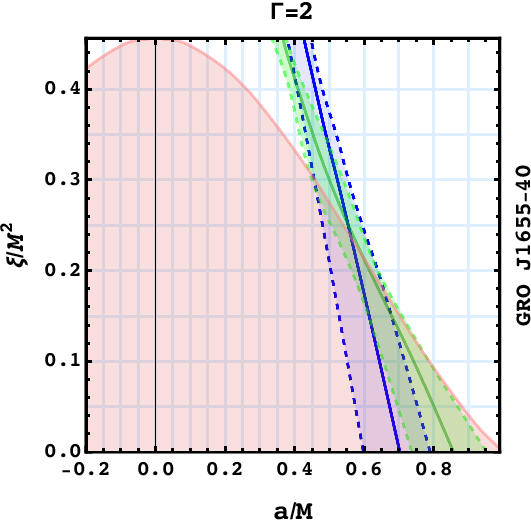}}\label{Fig:PE_5a}}
	\qquad
	\subfloat[\emph{}]{{\includegraphics[width=6cm]{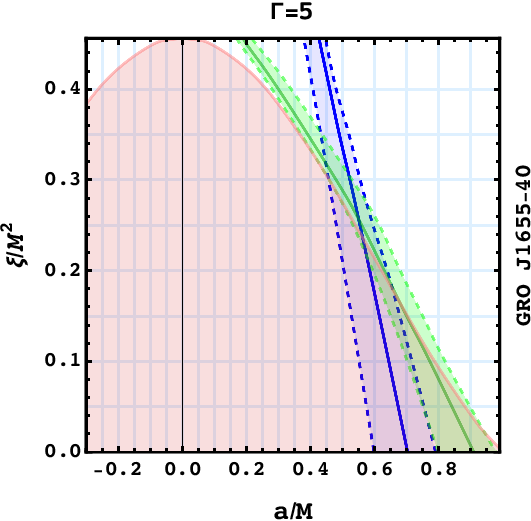}}\label{Fig:PE_5b}}
	\caption{\emph{Parameter constraints for GRO~J1655$-$40 have been depicted. The 'panel(a)' and 'panel (b)' correspond to Lorentz factors $\Gamma = 2$ and $\Gamma = 5$, respectively. In both cases, regular rotating black hole in ASG provides a consistent description of the observed radiative efficiency $\eta_{NT}$ and the jet power $P_{\mathrm{jet}}$. A more detailed explanation is provided in the main text.
	}}
	\label{Fig:Parameter_Estimation_5}
\end{figure}

For the present study, we adopt the spin parameter obtained from the Continuum Fitting method to compute the radiative efficiency $\eta_{NT}$. The corresponding region in the $\xi$ versus $a$ parameter space that satisfies the observed value of $\eta_{NT}$ within uncertainties is shown as the blue shaded area in ~\ref{Fig:Parameter_Estimation_5}.
Furthermore, we use the observed radio-flux density at $5$ GHz, measured as $2.42$ Jy, to estimate the kinetic power of the jet under the assumption of relativistic bulk motion with Lorentz factors $\Gamma = 2$ and $\Gamma = 5$. The corresponding jet powers are reported in ~\ref{Table2}, where an uncertainty of $0.3$ dex has been taken into account, consistent with standard assumptions~\cite{Narayan:2011eb,Middleton:2014cha}.

An examination of the shaded regions in ~\ref{Fig:Parameter_Estimation_5} reveals that the width of the blue zone, which corresponds to the parameter space consistent with the observed radiative efficiency, becomes narrower as the value of $\xi$ increases. This tightening of the boundaries indicates that higher values of $\xi$ tend to reduce the uncertainty around the theoretical estimate, effectively bringing the error margins closer to the central value of $\eta_{NT}$. 
In both panels, the green shaded areas represent the parameter ranges compatible with the measured jet power. As the Lorentz factor increases from $\Gamma = 2$ to $\Gamma = 5$, these green zones exhibit a modest shift toward higher values of the spin parameter. However, the overall structure of the allowed regions remains largely unchanged, suggesting that the spin dependence is only mildly sensitive to variations in the jet velocity.

A general trend becomes apparent: as the deformation parameter $\xi$ increases from zero to $0.4565$, the central value of the required spin parameter decreases from approximately $0.9$ to around $0.6$. This behavior implies that GRO~J1655$-$40 may be interpreted as a rapidly spinning black hole when no effect of deformation parameter is there, while it appears as a moderately rotating object within the framework of regular rotating  spacetime in ASG theory.
When both radiative efficiency and jet power are considered together, the comparison between the two Lorentz factor scenarios shows that the differences between the $\Gamma = 2$ and $\Gamma = 5$ cases are not substantial. For $\Gamma = 2$, the preferred combination of parameters is $a \simeq 0.55$ and $\xi \simeq 0.24$, while for $\Gamma = 5$, there is no crossing of two central value in black hole parameter space. However, for $\Gamma=5$, there is an overlap of blue and green region if the error bar is included within black hole parameter space. These results suggest a mild degeneracy between spin and the deformation parameter, which is further influenced by the assumed speed of the jet outflow.

\subsection{GRS 1915+105}

\begin{figure}[htbp!]
	\centering
	\subfloat[\emph{} ]{{\includegraphics[width=6cm]{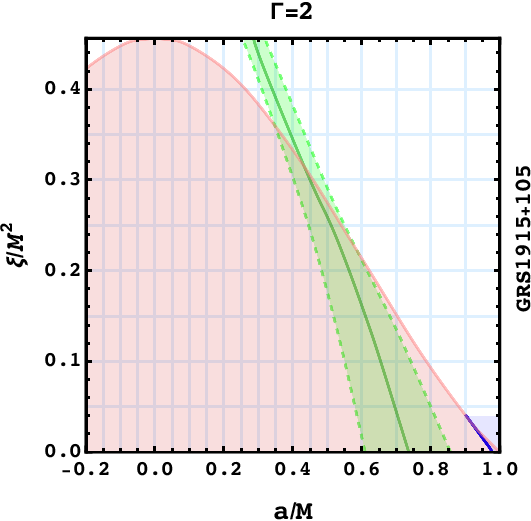}}\label{Fig:PE_6a}}
	\qquad
	\subfloat[\emph{}]{{\includegraphics[width=6cm]{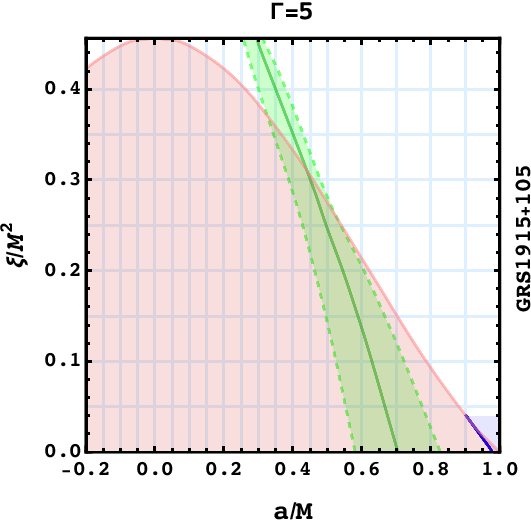}}\label{Fig:PE_6b}}
	\caption{\emph{Analysis of the source GRS~1915$+$105 has been shown. The results shown in the 'panel (a)' and 'panel (b)' demonstrate that the regular rotating spacetime in ASG faces difficulty in reconciling both the observed jet power and the radiative efficiency simultaneously. A detailed interpretation of this result is provided in the main body of the text.
	}}
	\label{Fig:Parameter_Estimation_6}
\end{figure}

GRS~1915$+$105 is a galactic X-ray binary system composed of a black hole and a K-type companion star, with an orbital period of approximately $34$ days~\cite{1994Natur.371...46M, Greiner:2001vb}. The mass of the black hole has been dynamically measured to be $M = 12.4^{+2.0}_{-1.8}\,M_\odot$~\cite{Reid:2014ywa}, while the source lies at a distance of $8.6^{+2.0}_{-1.6}$ kiloparsecs. The inclination angle of the orbital plane is estimated to be $60^\circ \pm 5^\circ$~\cite{Reid:2014ywa}.
The spin of the black hole has been inferred using the Continuum Fitting technique, which yields a value exceeding $a > 0.98$~\cite{McClintock:2006xd}. This high spin is adopted for the present study to compute the corresponding radiative efficiency. In ~\ref{Fig:Parameter_Estimation_6}, the blue shaded region enclosed by  solid curve represents the combinations of the deformation parameter $\xi$ and spin $a$ that are compatible with the observed value of $\eta_{NT}$.
The source is known to produce powerful radio jets, with a measured radio flux density of $0.912$ Jy at a frequency of $5$ GHz~\cite{1994Natur.371...46M}. Using this observed flux, the kinetic power of the jet has been estimated after applying Doppler corrections corresponding to bulk Lorentz factors $\Gamma = 2$ and $\Gamma = 5$. The resulting jet power values are listed in ~\ref{Table2}. In line with standard practice, an uncertainty of $0.3$ dex is assumed for each estimate.

In ~\ref{Fig:Parameter_Estimation_6}, the blue shaded region  represents the range of values for the deformation parameter $\xi$ and the spin $a$ that are consistent with the observed Novikov-Thorne radiative efficiency $\eta_{\mathrm{NT}}$. The figure indicates a clear trend: as the value of $\xi$ increases, a lower spin is required to match the efficiency derived from observations. From the plot, it is evident that the parameter $\xi$ cannot exceed a value of approximately $0.04$ if the model is to remain consistent with the measured radiative efficiency.
The green shaded region in Fig.~\ref{Fig:Parameter_Estimation_6} indicates the combinations of the spin parameter $a$ and the deviation parameter $\xi$ that are compatible with the observed jet power, considering the associated observational uncertainties. The solid and dashed green curves, as defined earlier, correspond to the central value and the bounds set by the assumed error. It is important to note that there is no overlap between the parameter ranges allowed by the jet power and those required to reproduce the radiative efficiency. This lack of intersection suggests that the model parameters cannot simultaneously account for both observables within the permitted error limits.

\section{Conclusion}\label{SEC:CONC}

In this study, we have explored the astrophysical implications of regular rotating black holes constructed within the framework of asymptotically safe gravity. These spacetimes incorporate ultraviolet modifications to the classical geometry through a running Newton's constant, leading to a quantum-corrected metric that remains regular throughout and preserves the asymptotic structure of the Kerr solution. We have shown that such corrections significantly alter the near-horizon geometry, particularly the location of the innermost stable circular orbit and the angular velocity of the event horizon. These two features govern the radiative efficiency of accretion disks and the power output of relativistic jets, respectively.

By examining a collection of well-observed black hole X-ray binaries, we have used radiative and jet-related measurements as complementary probes to constrain the spin and deformation parameters of the underlying quantum corrected metric. The analysis reveals that quantum corrections, even if small, can shift the predicted observables in measurable ways. In several cases, including A0620--00, GRO~J1655--40, and \\XTE~J1550--564, a physically allowed region in the spin -- deformation parameter space emerges that successfully explains both the disk efficiency and the observed jet power. This suggests that quantum gravitational effects near the horizon, encoded in the running of gravitational couplings, may already be playing a detectable role in shaping high-energy emissions from accreting black holes.

Importantly, the influence of the deformation parameter $\xi$ manifests differently across sources, depending on their spin, inclination, and jet velocity. Systems with moderate spins are particularly sensitive to these corrections, as the outward shift of the ISCO and the reduced frame-dragging allow lower-spin configurations to reproduce observational signatures typically attributed to rapidly rotating Kerr black holes. In contrast, high-spin systems like GRS~1915$+$105 pose a greater challenge, where the quantum-corrected model struggles to simultaneously match both observables within uncertainties. This mismatch may point to limitations of the current model or the need to include additional dynamical effects such as time-dependent accretion, magnetic flux saturation, or disc–jet coupling beyond the force-free limit.

Our results affirm that radiative efficiency and jet power are valuable observational tools for testing the strong-field structure of spacetime. Their sensitivity to the near-horizon geometry makes them ideal diagnostics for distinguishing classical black holes from regular ones predicted by quantum gravity. In particular, the Blandford–Znajek mechanism, when evaluated on quantum-corrected geometries, encodes the imprint of quantum deformation in the angular velocity of the horizon and in the electromagnetic energy extraction profile.

The interplay between observational constraints and quantum corrections studied here illustrates that astrophysical black holes are not merely endpoints of classical collapse but also laboratories for quantum gravity. As the precision of electromagnetic and gravitational wave measurements improves, models such as the one developed in this work will become increasingly relevant in evaluating the nature of gravity at its most extreme.

\iffalse
\begin{acknowledgements}
If you'd like to thank anyone, place your comments here
and remove the percent signs.
\end{acknowledgements}
\fi

\appendix

\section{Deriving Axisymmetric Rotating Spacetimes from Static Spherical Geometries: A Modified Approach to the Newman-Janis Algorithm}\label{App}

In this section, we outline a physically motivated framework to construct a rotating spacetime in Boyer-Lindquist coordinates, starting from a static and spherically symmetric seed metric. This procedure serves as a refined alternative to the conventional Newman-Janis algorithm (NJA), circumventing the ambiguities associated with complex coordinate transformations by instead relying on symmetry arguments and coordinate-dependent deformations of the metric functions.

We begin with a general static line element of the form:
\begin{equation}
	\dd s^2 = -G(r)\dd t^2 + \frac{\dd r^2}{F(r)} + H(r)(\dd\theta^2 + \sin^2\theta\,\dd\varphi^2).
\end{equation}
This geometry is assumed to be a solution to the Einstein field equations with the energy-momentum tensor given by
\begin{equation}
	T^{\mu}_{\;\nu} = \text{diag}(-\epsilon, p_r, p_\theta, p_\phi),
\end{equation}
describing a static anisotropic fluid distribution.

The aim is to obtain a consistent axisymmetric extension of this geometry that incorporates rotation while preserving the physical integrity of the source. Following the procedure developed in Ref.~\cite{Azreg-Ainou:2014pra}, the metric functions \( G(r), F(r), H(r) \) are promoted to more general functions $ A(r,\theta,a), B(r,\theta,a)$, $\Psi(r,\theta,a) $, where the spin parameter \( a \) is explicitly introduced as a deformation parameter. These generalized functions reduce to their static counterparts in the zero-spin limit:
\begin{equation}
	\lim_{a \to 0} A = G(r), \quad \lim_{a \to 0} B = F(r), \quad \lim_{a \to 0} \Psi = H(r).
\end{equation}

A key component of the construction is the definition of an auxiliary function:
\begin{equation}
	K(r) \equiv \sqrt{\frac{F(r)}{G(r)}}H(r),
\end{equation}
which facilitates the embedding of the rotational degrees of freedom into the static metric. With this, the modified rotating line element in Boyer-Lindquist coordinates takes the form
\begin{align}
	\dd s^2 = &-\frac{(F H + a^2\cos^2\theta)\Psi}{(K + a^2\cos^2\theta)^2}\dd t^2 + \frac{\Psi\,\dd r^2}{F H + a^2} + \Psi\,\dd \theta^2 \nonumber \\
	& - 2a\sin^2\theta\left[\frac{K - F H}{(K + a^2\cos^2\theta)^2}\right]\Psi\,\dd t\,\dd \phi \nonumber \\
	& + \Psi\sin^2\theta\left[1 + \frac{a^2\sin^2\theta(2K - F H + a^2\cos^2\theta)}{(K + a^2\cos^2\theta)^2}\right] \dd \phi^2.
\end{align}

The unknown function \( \Psi(r,\theta,a) \) is not arbitrary; its form must be such that the modified geometry remains a solution to Einstein's field equations. In the case of a rotating imperfect fluid, \( \Psi \) is constrained by two coupled partial differential equations:
\begin{align}
	&(K+a^2\cos^2\theta)^2 \left(3\Psi_{,r}\Psi_{,\theta} - 2\Psi \Psi_{,r\theta}\right) = 3a^2K_{,r}\Psi^2, \\
	&\left[K_{,r}^2 + K(2 - K_{,rr}) - a^2\cos^2\theta(2 + K_{,rr})\right]\Psi \nonumber \\
	&\quad + (K+a^2\cos^2\theta)\left[4\cos^2\theta \Psi_{,\theta} - K_{,r}\Psi_{,r}\right] = 0.
\end{align}

For a broad class of static solutions where the metric functions satisfy \( G = F \) and \( H = r^2 \) -- a structure that encompasses many regular black hole models -- a consistent solution to these equations is:
\begin{equation}
	\Psi = r^2 + a^2 \cos^2\theta.
\end{equation}
Inserting this into the general form yields a rotating geometry that closely resembles the Kerr metric, generalized through the function \( f(r) \), defined via the deviation from Schwarzschild geometry:
\begin{align}
	\dd s^2 &= -\left(1 - \frac{2f(r)}{\Sigma}\right)\dd t^2 + \frac{\Sigma}{\Delta} \dd r^2 + \Sigma \dd \theta^2 \nonumber \\
	&- \frac{4a f(r)\sin^2\theta}{\Sigma} \dd t\,\dd \phi + \frac{\mathcal{A}\sin^2\theta}{\Sigma} \dd \phi^2,
\end{align}
where the auxiliary functions are given by:
\begin{align}
	\Sigma &= r^2 + a^2 \cos^2\theta, \quad 2f(r) = r^2[1 - F(r)], \\
	\Delta &= r^2 F(r) + a^2, \quad \mathcal{A} = (r^2 + a^2)^2 - a^2\Delta \sin^2\theta.
\end{align}

To ensure the physical credibility of the rotating solution, it is essential to check its consistency with Einstein’s field equations. This is done by expressing the geometry in an orthonormal tetrad frame \( \{e^\mu_t, e^\mu_r, e^\mu_\theta, e^\mu_\phi\} \) adapted to the rotating spacetime. In this frame, the energy-momentum tensor takes the form:
\begin{equation}
	T^{\mu\nu} = \epsilon\, e^\mu_t e^\nu_t + p_r\, e^\mu_r e^\nu_r + p_\theta\, e^\mu_\theta e^\nu_\theta + p_\phi\, e^\mu_\phi e^\nu_\phi,
\end{equation}
where \( e^\mu_t \) represents the four-velocity of the rotating fluid. Direct computation confirms that the Einstein tensor components match this structure exactly, indicating that the solution is consistent with a rotating imperfect fluid configuration.

In summary, this method provides a robust and physically transparent pathway to construct rotating geometries from static spherically symmetric spacetimes. It preserves the causal and geometric structure of the seed solution and remains consistent with the field equations, offering a physically sound alternative to the traditional Newman-Janis prescription without relying on coordinate complexification.

\bibliographystyle{spphys}
\bibliography{references_GB}

\end{document}